\documentclass[a4paper,11pt]{article}
\usepackage{pos}
\usepackage{newtxtext,newtxmath}

\title{Elimination of QCD Renormalization Scale and Scheme Ambiguities}

\author*[a,b]{Leonardo Di Giustino}
\author[c]{Stanley J. Brodsky }
\author[a,b]{Philip G. Ratcliffe}
\author[d]{Xing-Gang Wu}
\author[e]{Sheng-Quan Wang}

\affiliation[a]{Department of Science and High Technology,
University of Insubria, via Valleggio 11, I-22100, Como, Italy\\}

\affiliation[b]{INFN, Sezione di Milano--Bicocca, 20126 Milano,
Italy\\}

\affiliation[c]{SLAC National Accelerator Laboratory, Stanford
University, Stanford, California 94039, USA\\}

\affiliation[d]{Department of Physics, Chongqing University,
Chongqing 401331, P.R. China\\}

\affiliation[e]{Department of Physics, Guizhou Minzu University,
Guiyang 550025, P.R. China\\}

\emailAdd{leonardo.digiustino@uninsubria.it}

\emailAdd{sjbth@slac.stanford.edu}

\emailAdd{philip.ratcliffe@uninsubria.it}

\emailAdd{wuxg@cqu.edu.cn}

\emailAdd{sqwang@alu.cqu.edu.cn}

\abstract{We present results for the thrust distribution in the
electron positron annihilation to the three jet process at NNLO in
the perturbative conformal window of QCD, as a function of the
number of flavors $N_f$. Given the existence of an infrared
interacting fixed point in this region, we can compare the
Conventional Scale Setting (CSS) and the Principle of Maximum
Conformality (PMC$_\infty$) methods along the entire
renormalization group flow from the highest energies to zero
energy. We then consider also the QED thrust, obtained as the
limit $N_c \rightarrow 0$ of the number of colors and we show
analogous comparison. QED in the low energy regime develops an
infrared non-interacting fixed point. Using these quantum field
theory limits as theoretical laboratories, we arrive at
interesting results showing new features of the PMC$_\infty$.}

\FullConference{16th International Symposium on Radiative Corrections: Applications of Quantum Field Theory to Phenomenology (
RADCOR2023)\\
28th May - 2nd June, 2023\\
Crieff, Scotland, UK\\}

\begin{document}
\maketitle

\section{Introduction}

The renormalization scale and scheme ambiguities are important
sources of errors, which greatly affect the results of the
Conventional Scale Setting (CSS) in perturbative QCD calculations.
Given the high precision reached by the experiments at LEP and
SLAC~\cite{ALEPH:2003obs,DELPHI:2003yqh,OPAL:2004wof,L3:2004cdh,SLD:1994idb}
in the measurement of the event shape variables and accuracy
achieved by higher order calculations from next-to-leading order
(NLO)
calculations~\cite{Ellis:1980wv,Kunszt:1980vt,Vermaseren:1980qz,Fabricius:1981sx,Giele:1991vf,Catani:1996jh}
to the next-to-next-to-leading
order(NNLO)~\cite{Gehrmann-DeRidder:2014hxk,Gehrmann-DeRidder:2007nzq,GehrmannDeRidder:2007hr,Weinzierl:2008iv,Weinzierl:2009ms}
and including resummation of the large
logarithms~\cite{Abbate:2010xh,Banfi:2014sua}, the elimination of
the uncertainties related to the scheme and scale ambiguities is
crucial in order to improve the results and to test the Standard
Model to the highest possible precision. In the particular case of
the three-jet event-shape distributions the conventional practice
of CSS leads to results which are in conflict with experimental
data and the extracted values of $\alpha_s$ deviate from the world
average~\cite{Workman:2022ynf}. A solution to the renormalization
scale ambiguity problem is provided by the {\emph{Principle of
Maximum Conformality}}
(PMC)~\cite{Brodsky:1982gc,Brodsky:2011ig,Brodsky:2011ta,Brodsky:2012rj,Mojaza:2012mf,Brodsky:2013vpa,Brodsky:2012ms,Wu:2014iba,Wang:2023ttk}.
This method provides a systematic way to eliminate renormalization
scheme-and-scale ambiguities from first principles by absorbing
the $\beta$ terms that govern the behavior of the running coupling
via the renormalization group equation. This procedure is
scheme-independent and leads to results free from divergent
renormalon terms~\cite{Beneke:1998ui}. The PMC procedure is
consistent with the standard Gell-Mann--Low method in the Abelian
limit, $N_c\rightarrow0$~\cite{Brodsky:1997jk} and can be
considered the non-Abelian extension of the
Serber--Uehling~\cite{Serber:1935ui,Uehling:1935uj} scale setting,
which is essential for precision tests of QED and atomic physics.
It should be emphasized that in a theory of unification of all
forces, electromagnetic, weak and strong interactions, such as the
Standard Model, or Grand Unification theories, one cannot simply
apply a different scale-setting or analytic procedure to different
sectors of the theory. The PMC offers the possibility to apply the
same method in all sectors of a theory, starting from first
principles, eliminating the renormalon growth, the scheme
dependence, scale ambiguities while satisfying the QED
Gell-Mann-Low scale-setting in the zero-color limit $N_c\to 0$. We
remark that PMC leads to scheme invariant results. We refer to
RS-scheme invariance as the invariance under the
extended-renormalization group and its equations (xRGE)\footnote{A
recent argument on PMC by Stevenson \cite{Stevenson:2023xvx},
based on the principle of minimum sensitivity (PMS), is incorrect.
Since the PMS is based on the assumption that all the unknown
higher-order terms give zero contribution to the pQCD series
\cite{Stevenson:1981vj}, its prediction directly breaks the
standard renormalization group invariance
\cite{Brodsky:2012ms,Wu:2014iba}, its pQCD series does not have
normal perturbative features \cite{Ma:2014oba} and can be treated
as an effective prediction only when we know the series up to high
enough orders and the conventional series has already shown good
perturbative features \cite{Ma:2017xef}. On the other hand, the
PMC respects all features of the renormalization group, and its
prediction satisfies all the requirements of standard
renormalization group invariance
\cite{Brodsky:2012ms,Wu:2014iba,Wu:2018cmb,Wu:2019mky}. A detailed
discussion on this point will be presented soon.}. Any other
relation among different quantities defined in different
approximations or obtained using different approaches, that can
have perturbative validity in QCD, can be improved using the PMC
and the residual dependence on the particular definition of the
"scheme" can be suppressed perturbatively by adding higher order
calculations. It can be shown that also in this case the results
obtained are scheme-independent (see Ref. \cite{Wu:2018cmb}). We
remark that in order to apply the PMC correctly, one should be
able to distinguish among the nature of the different $n_f$-terms,
whether they are related to the running of the coupling, to the
running of masses or to UV-finite diagrams and in a deeper
analysis also to the particular UV-divergent diagram (as discussed
in Refs. \cite{DiGiustino:2023jiq,DiGiustino:2022ggl}). Once all
$n_f$-terms have been associated with the correct diagram or
parameter, conformal coefficients are RG invariant and match the
coefficients of a conformal theory. Moreover, given that the PMC
preserves the RG invariance, it is possible to define CSR -
Commensurate Scale Relations\cite{Brodsky:2013vpa} among the
effective charges relating observables in different "schemes"
preserving all the group properties. Applying the PMC and the
CSRs, one can relate effective couplings, as also conformal
coefficients, leading to scheme independent results for the
observables. Applications of the PMC to different quantities (see
Ref. \cite{Huang:2021hzr,Wang:2020ckr}) have recently shown a
direct improvement of theoretical predictions. The recently
developed {\emph{Infinite-Order Scale-Setting using the Principle
of Maximum Conformality}} (PMC$_\infty$) has been shown to
significantly reduce the theoretical errors in Event Shape
Variable distributions, highly improving also the fit with the
experimental
data\cite{DiGiustino:2023jiq,Wang:2021tak,DiGiustino:2020fbk,DiGiustino:2021nep}.
An improved theoretical prediction on $\alpha_s$ with respect to
the world average has also been shown in
Refs.\cite{Wang:2019ljl,Wang:2019isi}. In this article we consider
the thrust distribution in the region of flavors and colors near
the upper bound of the conformal window, i.e. $N_f\sim 11/2 N_c$ ,
where the IR fixed point can be reliably accessed in perturbation
theory and we compare the two renormalization scale setting
methods, the CSS and the PMC$_\infty$. In this region we are able
to deduce the full solution at NNLO in the strong coupling. We
also compare the two methods in the QED, $N_c\rightarrow 0$, limit
of thrust.

\section{Two-loop solution and the perturbative conformal window \label{twoloops}}

The strong coupling dependence on the scale can be described
introducing the $\beta$-function given by:
 \begin{equation}
\frac{1}{4 \pi}\frac{d\alpha_s(Q^2)}{d \log Q^2}=\beta(\alpha_s),
\label{betafun1} \end{equation}  and \begin{equation}
\beta\left(\alpha_{s}\right)=-\left(\frac{\alpha_{s}}{4
\pi}\right)^{2} \sum_{n=0} \left(\frac{\alpha_{s}}{4
\pi}\right)^{n} \beta_{n}. \label{betafun10}\end{equation}
Neglecting quark masses, the first two $\beta$-terms are RS
independent and they have been calculated in
Refs.~\cite{Gross:1973id,Politzer:1973fx,Caswell:1974gg,Jones:1974mm,Egorian:1978zx}
for the $\overline{\rm MS}$ scheme:
 \begin{eqnarray}\beta_{0}&=& \frac{11}{3}C_{A} -\frac{4}{3}T_{R}N_{f},\\
\beta_{1}&=&
\frac{34}{3}C_{A}^{2}-4\left(\frac{5}{3}C_{A}+C_{F}\right)T_{R}N_{f},\end{eqnarray}
where $C_F=\frac{\left(N_{c}^{2}-1\right)}{2 N_{c}}$, $C_A=N_c$
and $T_R=1/2$ are the color factors for the ${\rm SU(3)}$ gauge group~\cite{Mojaza:2010cm}. \\
In order to determine the solution for the strong coupling
$\alpha_s$ at NNLO, it is convenient to introduce the following
notation: $x(\mu)\equiv \frac{\alpha_s(\mu)}{2 \pi}$,
$t=\log(\mu^2/\mu_0^2)$, $B=\frac{1}{2}\beta_0$ and
$C=\frac{1}{2}\frac{\beta_1}{\beta_0}$, $x^*\equiv -\frac{1}{C}$.
The truncated NNLO approximation of the Eq.~\ref{betafun1} leads
to the differential equation:
\begin{equation}
\frac{dx}{dt}=-B x^2(1+C x). \label{lambert1}
\end{equation}
An implicit solution of Eq.~\ref{lambert1} is given by the Lambert
$W(z)$ function:
\begin{equation}
W e^W = z, \label{W}
\end{equation}
with: $ W=\left(\frac{x^*}{ x}-1\right)$. The general solution for
the coupling is:
\begin{eqnarray}
x &=& \frac{x^*}{1+W} , \\
 z &=& e^{\frac{x^*}{x_0}-1} \left(\frac{x^*}{x_0}-1 \right) \left( \frac{\mu^2}{\mu_0^2}
\right)^{x^* B}. \label{xz}
\end{eqnarray}
We shall discuss here the solutions to Eq.~\ref{lambert1} with
respect to the particular initial phenomenological value
$x_0\equiv \alpha_s(M_Z) /(2\pi)= 0.01876 \pm 0.00016$ given by
the coupling determined at the $Z^0$ mass
scale~\cite{Workman:2022ynf}. This value introduces an upper
boundary on the number of flavors: $\bar{N}_f =x^{*-1}(x_0)=
15.222 \pm 0.009$, which narrows the range for the IR fixed point
discussed by Banks and Zaks~\cite{Banks:1981nn}. The only
physically accessible range is $N_f< \bar{N}_f$, since for $N_f>
\bar{N}_f$ we no longer have the asymptotically free UV behavior.
In the range $N_f<\bar{N_f^1}$ we have $B>0$, $C>0$ and the
physical solution is given by the $W_{-1}$ branch, while for
$\bar{N_f^1}< N_f < \bar{N_f}$ the solution for the strong
coupling is given by the $W_{0}$ branch. Where the values
$\bar{N_f^0}=\frac{11}{2}N_c$, $\bar{N_f^1}=\frac{34 N_c^3}{13
N_c^2-3}$,  with $\bar{N_f^0}> \bar{N_f^1}$,  are the zeros of
$\beta_0,\beta_1$ respectively. The two-dimensional region in the
number of flavors and colors where asymptotically free QCD
develops an IR interacting fixed point is colloquially known as
the {\it conformal window of pQCD}. Different solutions for the
NNLO equation can be achieved using different schemes, i.e.
different definitions of the $\Lambda$ scale parameter, as shown
in Ref.~\cite{Gardi:1998qr}. In general, IR and UV fixed points of
the $\beta$-function can also be determined at different values of
the number of colors $N_c$ (different gauge group $SU(N)$) and
$N_f$ extending this analysis also to other gauge
theories~\cite{Ryttov:2017khg}. An extension of the perturbative
QCD coupling and its $\beta$-function in the IR region can be
found in Ref. \cite{Brodsky:2010ur}.

\section{The thrust distribution according to $N_f$}

The thrust distribution and the event shape variables are a
fundamental tool in order to probe the geometrical structure of a
given process at colliders and for the measurement of the strong
coupling $\alpha_s$~\cite{Kluth:2006bw}. Thrust ($T$) is defined
as
\begin{eqnarray}
T=\max\limits_{\vec{n}}\left(\frac{\sum_{i}|\vec{p}_i\cdot\vec{n}|}{\sum_{i}|\vec{p}_i|}\right),
\end{eqnarray}
where the sum runs over all particles in the hadronic final state,
and $\vec{p}_i$ denotes the three-momentum of particle $i$. The
unit vector $\vec{n}$ is varied to maximize thrust $T$, and the
corresponding $\vec{n}$ is called the thrust axis and denoted by
$\vec{n}_T$. The variable $(1-T)$ is often used, which for the LO
of 3-jet production is restricted to the range $(0<1-T<1/3)$. We
have a back-to-back or a spherically symmetric event respectively
at $T=1$ and at $T=2/3$ respectively.

\begin{figure}[h]
 \centering
\hspace*{-0.5cm}
\includegraphics[width=12cm]{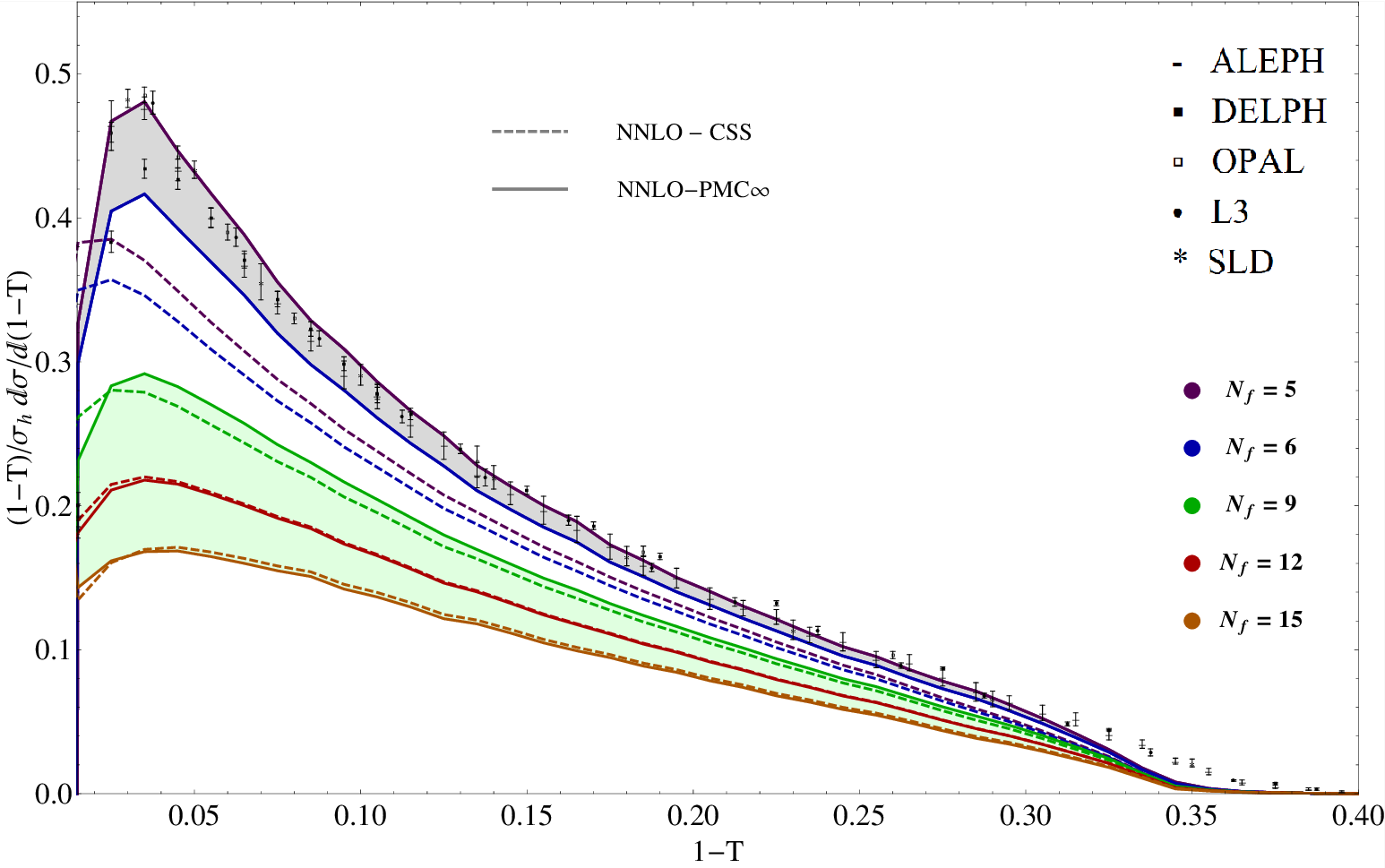}
\caption{Thrust distributions for different values of $N_f$, using
the PMC$_\infty$ (solid line) and the CSS (dashed
line)~\cite{DiGiustino:2021nep}. The Green shaded area is the
results for the values of $N_f$ taken in the conformal window. The
experimental data points are taken from the ALEPH, DELPHI,OPAL,
L3, SLD
experiments~\cite{ALEPH:2003obs,DELPHI:2003yqh,OPAL:2004wof,L3:2004cdh,SLD:1994idb}.}
\label{confthrust}
\end{figure}

In general, a normalized IR-safe single-variable observable, such
as the thrust distribution for the $e^+ e^-\rightarrow
3jets$~\cite{DelDuca:2016ily,DelDuca:2016csb}, is the sum of pQCD
contributions calculated up to NNLO at the initial renormalization
scale $\mu_0=\sqrt{s}=M_{Z}$:
\begin{eqnarray}
\frac{1}{\sigma_{tot}} \! \frac{O d \sigma(\mu_{0})}{d O}\! & = &
\left\{ x_0 \cdot \frac{ O d \overline{A}_{\mathit{O}}(\mu_0)}{d
O} +
x_0^2 \cdot \frac{ O d \overline{B}_{\mathit{O}}(\mu_0)}{d O} \right. \nonumber  \\
 & & + \left. x_0^{3} \cdot \frac{O d\overline{C}_{\mathit{O}}(\mu_0)}{d
O}+ {\cal O}(\alpha_{s}^4) \right\}
 \label{observable1-thrust}
\end{eqnarray}
where $x(\mu)\equiv \alpha_s(\mu)/(2\pi)$, $O$ is the selected
event shape variable, $\sigma$ the cross section of the process,
\begin{equation}\sigma_{tot}=\sigma_{0} \left( 1+x_0 A_{t o t}+ x_0^{2} B_{t
o t}+ {\cal O}\left(\alpha_{s}^{3}\right)\right),\end{equation} is
the total hadronic cross section and $\overline{A}_O,
\overline{B}_O, \overline{C}_O$ are respectively the normalized
LO, NLO and NNLO coefficients:
\begin{eqnarray}
\overline{A}_{O} &=&A_{O}, \nonumber \\
\overline{B}_{O} &=&B_{O}-A_{t o t} A_{O}, \\
\overline{C}_{O} &=&C_{O}-A_{t o t} B_{O}-\left(B_{t o t}-A_{t o
t}^{2}\right) A_{O}, \nonumber
\end{eqnarray}
where $A_O, B_O, C_O$ are the coefficients normalized to the
tree-level cross section $\sigma_0$ calculated by MonteCarlo (see
e.g. the EERAD and Event2
codes~\cite{Gehrmann-DeRidder:2014hxk,Gehrmann-DeRidder:2007nzq,GehrmannDeRidder:2007hr,Weinzierl:2008iv,Weinzierl:2009ms})
and $A_{\mathit{tot}}, B_{\mathit{tot}}$ are:
\begin{eqnarray}
A_{\mathit{tot}} &= & \frac{3}{2} C_F,  \nonumber \\
B_{\mathit{tot}} &= & \frac{C_F}{4}N_c +\frac{3}{4}C_F
\frac{\beta_0}{2} (11-8\zeta(3)) -\frac{3}{8} C_F^2,
 \label{norm}
\end{eqnarray}
where $\zeta$ is the Riemann zeta function. In the case of
conventional scale setting, the renormalization scale is set to
$\mu_0=\sqrt{s}=M_Z$ and theoretical uncertainties are evaluated
using standard criteria. In this case, we have used the definition
given in Ref.~\cite{Gehrmann-DeRidder:2007nzq} of the parameter
$\delta$; we define the average error for the event shape variable
distributions as:
 \begin{equation}
\bar{\delta}=\frac{1}{N} \sum_i^N \frac{ {\rm
max}_{\mu}(\sigma_i(\mu))- {\rm min}_{\mu} (\sigma_i(\mu))}{2
\sigma_i(\mu=M_Z)}, \label{delta}
\end{equation} where $i$ is the index of the bin and $N$ is the
total number of bins, the renormalization scale is varied in the
range: $\mu \in [M_Z/2; 2 M_Z]$. We use here the two-loop solution
for the running coupling $\alpha_s(Q)$. According to the
PMC$_\infty$ (for a detailed analysis see
Ref.~\cite{DiGiustino:2023jiq,DiGiustino:2020fbk,DiGiustino:2021nep}),
Eq.~\ref{observable1-thrust} becomes:
\begin{equation}
\frac{1}{\sigma_{tot}} \! \frac{O d \sigma(\mu_{\rm
I},\tilde{\mu}_{\rm II},\mu_{0})}{d O}=
\left\{\overline{\sigma}_{\rm I}+\overline{\sigma}_{\rm
II}+\overline{\sigma}_{\rm III}+ {\cal O}(\alpha_{s}^4)
\right\},\label{observable3}
\end{equation}
where the $\overline{\sigma}_{N}$ are normalized subsets given by:
\begin{eqnarray}
\overline{\sigma}_{\rm I} &=& A_{\mathit{Conf}} \cdot x_{\rm I}, \nonumber  \\
\overline{\sigma}_{\rm II} &=& \left( B_{\mathit{Conf}}+\eta
A_{\mathit{tot}} A_{\mathit{Conf}} \right)\cdot x_{\rm II}^2
 - \eta A_{\textrm{tot}} A_{\mathit{Conf}} \cdot x_0^2 \nonumber \\
 & & \hspace{2.3cm} -A_{\textrm{tot}} A_{\mathit{Conf}}\cdot x_0 x_{\rm I},  \nonumber \\
\overline{\sigma}_{\rm III} &=&\!\! \left( C_{\mathit{Conf}} \!-\!
A_{\textrm{tot}} \!B_{\mathit{Conf}}\!-\!(B_{\textrm
{tot}}-A_{\textrm{tot}}^{2}) A_{\mathit{Conf}}\right) \cdot x_0^3,
\label{normalizedcoeff}
\end{eqnarray}
where $A_{\mathit{Conf}}, B_{\mathit{Conf}}, C_{\mathit{Conf}}$
are the scale-invariant conformal coefficients (i.e. the
coefficients of each perturbative order not depending on the scale
$\mu_R$) while $x_{\rm I},x_{\rm II},x_0$ are the couplings
determined at the PMC$_\infty$ scales: $\mu_{\rm
I},\tilde{\mu}_{\rm II}, M_Z$ respectively. The PMC$_\infty$
scales are given by:
\begin{eqnarray}
\large{\mu}_{\rm I} & = & \sqrt{s} \cdot e^{f_{sc}-\frac{1}{2} B_{\beta_0}}\hspace{3.5cm}{ \scriptstyle (1-T)<0.33},  \label{icfscale1}  \\
\large{\tilde{\mu}}_{\rm II} & =& \left\{
\begin{array}{lr} \sqrt{s} \cdot e^{f_{sc}-\frac{1}{2} C_{\beta_0}
\cdot \frac{B_{\mathit{Conf}}}{B_{\mathit{Conf}}+\eta \cdot
A_{\mathit{tot}} A_{\mathit{Conf}} }} &   {\scriptstyle (1-T)<0.33},  \\
  \sqrt{s}\cdot e^{f_{sc}-\frac{1}{2} C_{\beta_0}}, & { \scriptstyle
  (1-T)>0.33},\\
  \label{icfscale2}
 \end{array} \right.
 \label{PMC12}
\end{eqnarray}
where the $f_{sc}$ is a general scheme factor which is
$f_{sc}\equiv 0$ for the QCD case in $\overline{\rm MS}$-scheme.
Normalized subsets for the region $(1-T)>0.33$ can be simply
achieved by setting $A_{\mathit{Conf}}\equiv 0$ in the
Eq.~\ref{normalizedcoeff}. Results for the thrust distribution
calculated using the  NNLO solution for the coupling
$\alpha_s(\mu)$, at different values of the number of flavors,
$N_f$,  is shown in Fig.~\ref{confthrust}.

\begin{figure}[ht]
\centering
\includegraphics[width=0.7\textwidth]{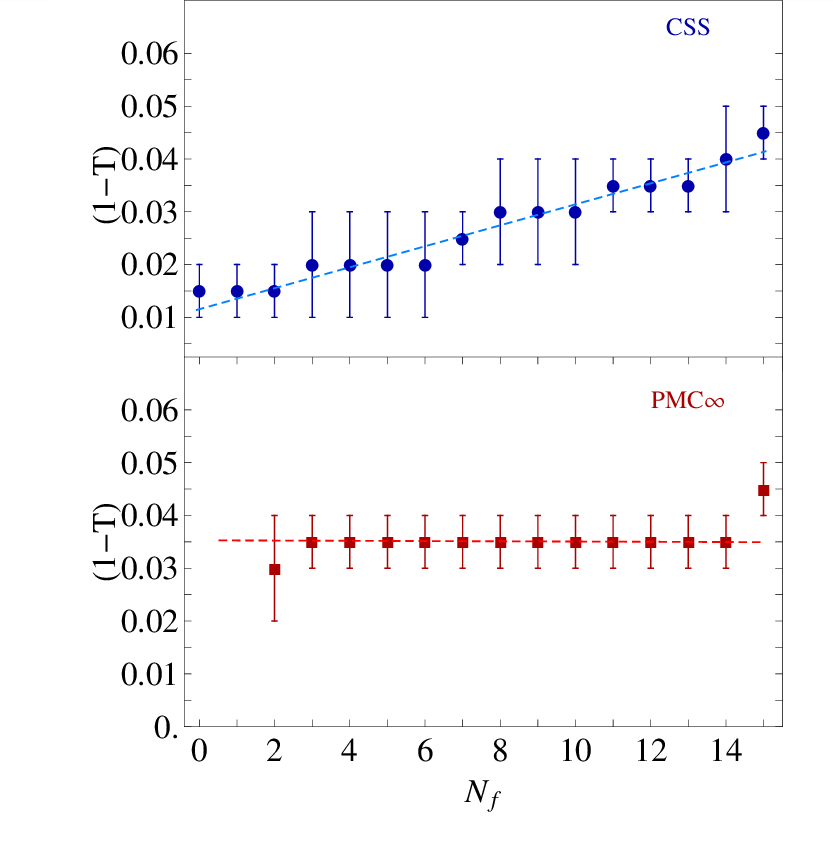}
\caption{Comparison of the position of the peak for the thrust
distribution using the CSS and the PMC$_\infty$ vs the number of
flavors, $N_f$. Dashed lines indicate the particular trend in each
graph~\cite{DiGiustino:2021nep}.} \label{peaks}
\end{figure}

\begin{figure}[htb]
\centering
\includegraphics[width=10cm]{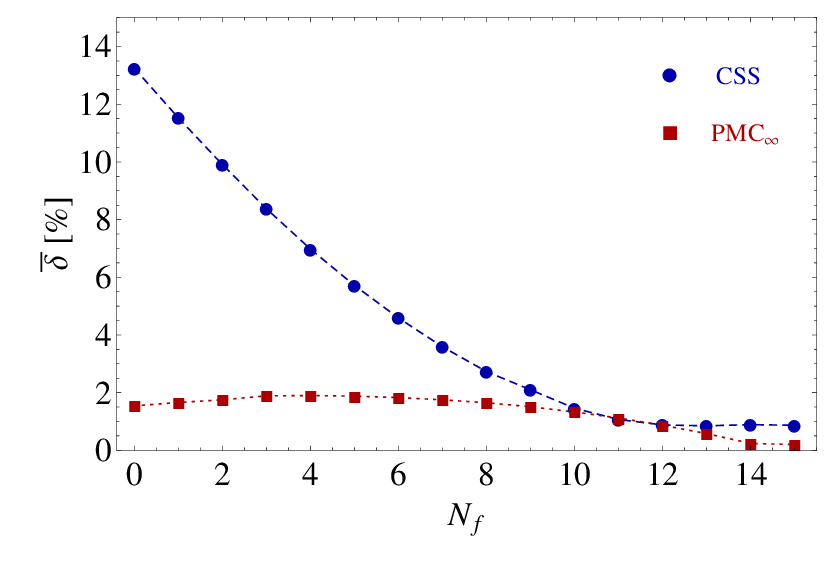}
\caption{Comparison of the average theoretical error,
$\bar{\delta}$, calculated using standard criteria in the range:
$0.005<(1-T)<0.4$, using the CSS and the PMC$_\infty$ for the
thrust distribution vs the number of flavors,
$N_f$~\cite{DiGiustino:2021nep}.} \label{err}
\end{figure}

A direct comparison between PMC$_\infty$ (solid line) and CSS
(dashed line) is shown at different values of the number of
flavors. We notice that, despite the phase transition (i.e. the
transition from an infrared finite coupling to an infrared
divergent coupling), the curves given by the PMC$_\infty$ at
different $N_f$, preserve with continuity the same characteristics
of the conformal distribution setting $N_f$ out of the conformal
window of pQCD. We notice that by decreasing $N_f$, the peak
increases and this is mainly due to larger values of the
$\beta_0,\beta_1$ coefficients, obtaining a better match with the
data for values in the range $5\leq N_f\leq 6$. The position of
the peak of the thrust distribution is well preserved varying
$N_f$ in and out of the conformal window using the PMC$_\infty$,
while there is constant shift towards lower values using the CSS.
These trends are shown in Fig.~\ref{peaks}. We notice that in the
central range, $2<N_f<15$, the position of the peak is exactly
preserved using the PMC$_\infty$ and overlaps with the position of
the peak shown by the experimental data. Theoretical uncertainties
on the position of the peak have been calculated using standard
criteria, i.e. varying the remaining initial scale value in the
range $M_Z/2 \leq \mu_0 \leq 2 M_Z$, and considering the lowest
uncertainty given by the half of the spacing between two adjacent
bins.
%
Using the definition given in Eq.~\ref{delta}, we have determined
the average error, $\bar{\delta}$, calculated in the interval
$0.005<(1-T)<0.4$ of thrust and results for CSS and PMC$_\infty$
are shown in Fig.~\ref{err}. We notice that the PMC$_\infty$ in
the perturbative and IR conformal window, i.e. $12<N_f<\bar{N}_f$,
which is the region where $\alpha_s(\mu)<1$ in the whole range of
the renormalization scale values, from $0$ up to $\infty$, the
average error given by PMC$_\infty$ tends to zero ($\sim
0.23-0.26\%$) while the error given by the CSS tends to remain
constant ($0.85-0.89\%$). The comparison of the two methods shows
that, out of the conformal window, $N_f<\frac{34 N_c^3}{13
N_c^2-3}$, the PMC$_\infty$ leads to a higher precision.

\section{The thrust distribution in the Abelian limit
$N_c\rightarrow 0$ \label{sec:qedthrust}}

\begin{figure}[htb]
\centering
\includegraphics[width=12cm]{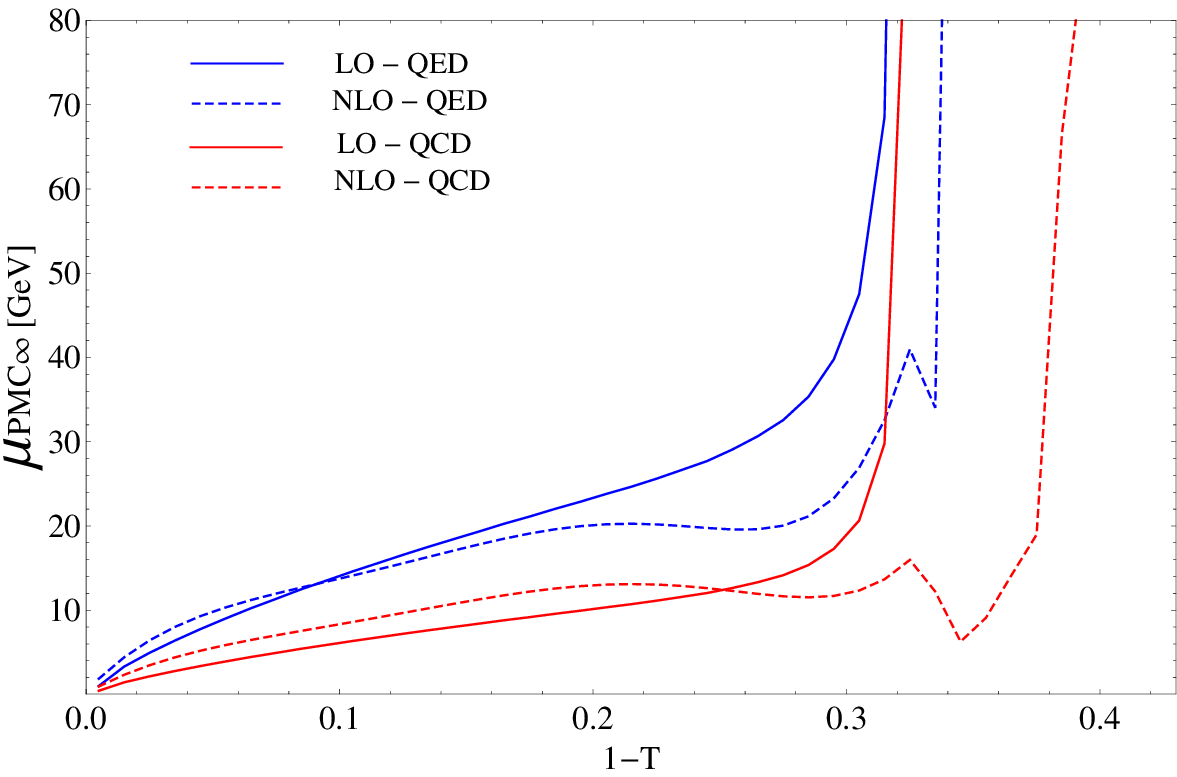}
\caption{PMC$_\infty$ scales for the thrust distribution: LO-QCD
scale (solid red); LO-QED scale (solid blue);NLO-QCD scale (dashed
red); NLO-QED scale (dashed blue)~\cite{DiGiustino:2021nep}.}
\label{scales}
\end{figure}

\begin{figure}[htb]
\centering \vspace{1cm}
\includegraphics[width=12cm]{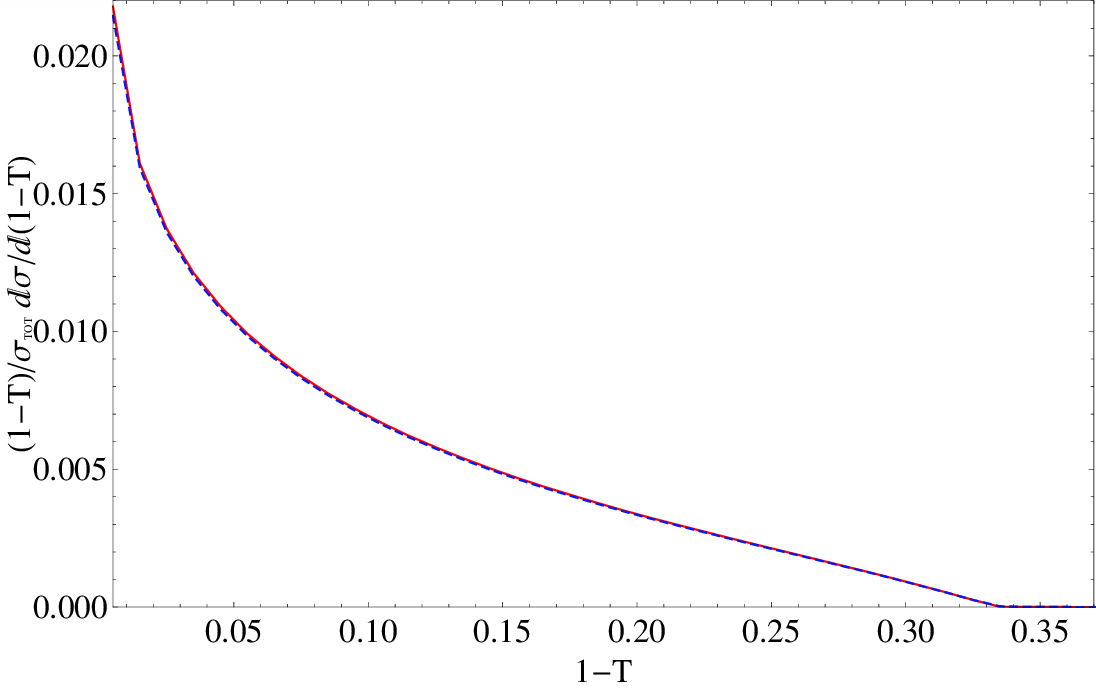}
\caption{Thrust distributions in the QED limit at NNLO using the
PMC$_\infty$ (solid red) and the CSS (dashed
blue)~\cite{DiGiustino:2021nep}.} \label{qedthrust}
\end{figure}

We consider now the thrust distribution in U(1) Abelian QED, which
rather than being infrared interacting is infrared free. We obtain
the QED thrust distribution performing the $N_c\rightarrow 0$
limit of the QCD thrust at NNLO according
to~\cite{Brodsky:1997jk,Kataev:2015yha}. In the zero number of
colors limit the gauge group color factors are fixed by $N_A=1,$
$C_F=1,$ $T_R=1,$ $C_A=0,$ $N_c=0,$ $N_f=N_l$, where $N_l$ is the
number of active leptons, while the $\beta$-terms and the coupling
rescale as $\beta_n/C_F^{n+1}$ and $\alpha_s \cdot C_F$
respectively. In particular $\beta_0=-\frac{4}{3}N_l$ and
$\beta_1=-4 N_l$ using the normalization of Eq.~\ref{betafun1}.
According to this rescaling of the color factors we have
determined the QED thrust and the QED PMC$_\infty$ scales. For the
QED coupling, we have used the analytic formula for the effective
fine structure constant in the $\overline{\textrm{MS}}$-scheme:
\begin{equation}
{\alpha(Q^2)} = {\alpha \over  { \left(1 -\Re e
\Pi^{\overline{\textrm{MS}}} (Q^2)\right)}},
\end{equation}
with $\alpha^{-1}\equiv \alpha(0)^{-1}= 137.036$ and the vacuum
polarization function ($\Pi$) calculated perturbatively at two
loops including contributions from leptons, quarks and $W$ boson.
The QED PMC$_\infty$ scales have the same form of
Eqs.~\ref{icfscale1} and \ref{icfscale2} with the factor for the
$\overline{\textrm{MS}}$-scheme set to $f_{sc}\equiv 5/6$ and the
$\eta$ regularization parameter introduced to cancel singularities
in the NLO PMC$_\infty$ scale $\mu_{\rm II}$ in the $N_c
\rightarrow 0$ limit tends to the same QCD value, $ \eta=3.51 $. A
direct comparison between QED and QCD PMC$_\infty$ scales is shown
in Fig.~\ref{scales}.
We note that in the QED limit the PMC$_\infty$ scales have
analogous dynamical behavior as those calculated in QCD,
differences arise mainly owing to the $\overline{\textrm{MS}}$
scheme factor reabsorption, the effects of the $N_c$ number of
colors at NLO are also negligible. Thus we notice that perfect
consistency is shown from QCD to QED using the PMC$_\infty$
method. The normalized QED thrust distribution is shown in
Fig.~\ref{qedthrust}. We note that the curve is peaked at the
origin, $T=1$, which suggests that the three-jet event in QED
occurs with a rather back-to-back symmetry. Results for the CSS
and the PMC$_\infty$ methods in QED are of the order of
$O(\alpha)$ and show very small differences, given the good
convergence of the theory.

\section{Conclusion}

We have investigated for the first time the thrust distribution
within the perturbative conformal window of QCD and in QED and we
have compared results obtained using both the CSS and the
PMC$_\infty$ scale setting. In the latter case the results are in
perfect agreement with the Gell-Man--Low scheme. Results for
different values of the $N_f$ factor show that the PMC$_\infty$
scale setting leads to higher precision and are in agreement with
the data in a wide range of the selected event shape variable.
Moreover the thrust distributions in the conformal window have
similar shapes to those of the physical values of $N_f$ and the
position of the peak is preserved when one applies the
PMC$_\infty$ method. Thus, even though the peak is a property
directly related to the resummation of the large-logarithms in the
low-energy
region~\cite{Catani:1991kz,Catani:1992ua,Catani:1996yz,Aglietti:2006wh,Aglietti:2007bp,Abbate:2010xh},
the correct position of the peak can be considered in fact a
conformal property and it can be related to the use of the
PMC$_\infty$ scales.

\section*{Acknowledgements}
LDG thanks the organizers of RADCOR 2023 for the opportunity to
give his presentation. This research was supported in part by the
Department of Energy contract {\text DE-AC02-76SF00515} (SJB).
SLAC-PUB-17750.

\end{document}